\documentclass[conference]{IEEEtran} 

\IEEEoverridecommandlockouts

\usepackage{cite}
\usepackage{amsmath,amssymb,amsfonts}
\usepackage{algorithmic}
\usepackage{graphicx}
\usepackage{textcomp}
\usepackage{xcolor}
\def\BibTeX{{\rm B\kern-.05em{\sc i\kern-.025em b}\kern-.08em
    T\kern-.1667em\lower.7ex\hbox{E}\kern-.125emX}}

\usepackage{multirow}
\usepackage{url}
\usepackage{breakurl}
\usepackage{enumitem}
\usepackage{verbatim}
\usepackage{xspace}
\usepackage{diagbox}
\usepackage{array}
\usepackage{cite}

\usepackage{breakurl}
\usepackage[breaklinks]{hyperref}

\usepackage{tcolorbox}
\newcommand{\boxmargin}{0mm}
\newcommand{\figmargin}{-5pt}

\newcommand{\myauthornote}[3]{{}}

\newcommand{\CommitgenData}{CommitGen\textsubscript{data}\xspace}
\newcommand{\NngenData}{NNGen\textsubscript{data}\xspace}
\newcommand{\CodisumData}{CoDiSum\textsubscript{data}\xspace}
\newcommand{\MlData}{MultiLang\textsubscript{data}\xspace}
\newcommand{\PtrgenData}{PtrGen\textsubscript{data}\xspace}
\newcommand{\AtomData}{ATOM\textsubscript{data}\xspace}
\newcommand{\mcmd}{MCMD\xspace}
\newcommand{\mcmdjava}{MCMD\textsubscript{Java}\xspace}
\newcommand{\mcmdcsharp}{MCMD\textsubscript{C\#}\xspace}
\newcommand{\mcmdcpp}{MCMD\textsubscript{C++}\xspace}
\newcommand{\mcmdpython}{MCMD\textsubscript{Py}\xspace}
\newcommand{\mcmdjs}{MCMD\textsubscript{JS}\xspace}

\newcommand{\Commitgen}{CommitGen\xspace}
\newcommand{\Nmt}{NMT\xspace}
\newcommand{\Codisum}{CoDiSum\xspace}
\newcommand{\Nngen}{NNGen\xspace}
\newcommand{\CCVec}{CC2Vec\xspace}
\newcommand{\ChangeDoc}{ChangeDoc\xspace}
\newcommand{\Ptrnet}{PtrGNCMsg\xspace}

\newcommand{\Atom}{ATOM\xspace}

\newcommand{\bleumoses}{B-Moses\xspace}
\newcommand{\bleumosesnorm}{B-Norm\xspace}
\newcommand{\bleunltksmfive}{B-CC\xspace}

\newcommand{\Fig}{Figure\xspace}
\newcommand{\Tab}{Table\xspace}
\newcommand{\Sec}{Section\xspace}

\usepackage{comment}
\usepackage{booktabs}

\begin{document}

\title{On the Evaluation of Commit Message Generation Models: An Experimental Study}

\author{\IEEEauthorblockN{
Wei Tao\IEEEauthorrefmark{2}\IEEEauthorrefmark{6}\thanks{\IEEEauthorrefmark{6}Work is done during internship at Microsoft Research Asia.}, 
Yanlin Wang\IEEEauthorrefmark{3}\IEEEauthorrefmark{1}\thanks{\IEEEauthorrefmark{1}Yanlin Wang is the corresponding author.}, 
Ensheng Shi\IEEEauthorrefmark{4}\IEEEauthorrefmark{6}, 
Lun Du\IEEEauthorrefmark{3}, 
Shi Han\IEEEauthorrefmark{3}, \\
Hongyu Zhang\IEEEauthorrefmark{5}, 
Dongmei Zhang\IEEEauthorrefmark{3} and 
Wenqiang Zhang\IEEEauthorrefmark{2}
\IEEEauthorblockA{\IEEEauthorrefmark{2}Academy for Engineering and Technology, Fudan University, \{wtao18, wqzhang\}@fudan.edu.cn}
\IEEEauthorblockA{\IEEEauthorrefmark{3}Microsoft Research Asia, \{yanlwang, lun.du, shihan, dongmeiz\}@microsoft.com}
\IEEEauthorblockA{\IEEEauthorrefmark{4}Xi'an Jiaotong University, s1530129650@stu.xjtu.edu.cn} 
\IEEEauthorblockA{\IEEEauthorrefmark{5}The University of Newcastle, hongyu.zhang@newcastle.edu.au}
}  \\\vspace{-30pt}}
\maketitle


\tcbset{colback=gray!8,
        colframe=black,
        width=9cm,
        arc=2mm, auto outer arc,
        boxrule = 0.7pt,
        left = \boxmargin, right = \boxmargin, top = \boxmargin, bottom = \boxmargin,
        leftright skip=0.0mm
}

\begin{abstract}
Commit messages are natural language descriptions of code changes, which are important for program understanding and maintenance. 
However, writing commit messages manually is time-consuming and laborious, especially when the code is updated frequently. 
Various approaches utilizing generation or retrieval techniques have been proposed to automatically generate commit messages. 
To achieve a better understanding of how the existing approaches perform in solving this problem, this paper conducts a systematic and in-depth analysis of the state-of-the-art models and datasets. We find that:  
(1) Different variants of the BLEU metric are used in previous works, which affects the evaluation and understanding of existing methods. 
(2) Most existing datasets are crawled only from Java repositories while repositories in other programming languages are not sufficiently explored. 
(3) Dataset splitting strategies can influence the performance of existing models by a large margin. Some models show better performance when the datasets are split by commit, while other models perform better when the datasets are split by timestamp or by project. 
Based on our findings, we conduct a human evaluation and find the BLEU metric that best correlates with the human scores for the task. 
We also collect a large-scale, information-rich, and multi-language commit message dataset MCMD and evaluate existing models on this dataset. 
Furthermore, we conduct extensive experiments under different dataset splitting strategies and suggest the suitable models under different scenarios. 
Based on the experimental results and findings, we provide feasible suggestions for comprehensively evaluating commit message generation models and discuss possible future research directions. We believe this work can help practitioners and researchers better evaluate and select models for automatic commit message generation. Our source code and data are
available at \url{https://github.com/DeepSoftwareAnalytics/CommitMsgEmpirical}.

\end{abstract}


\vspace{-10pt}
\section{Introduction}
\vspace{-2pt}

A tremendous amount of code is generated and updated every day, which is often maintained by version control systems such as Git. In the course of software development and maintenance, developers could frequently change their code to fix bugs, add features, perform refactoring, etc. Commits keep track of these code changes. Each commit is associated with a message which describes what and why these code changes are made. 
Commit messages can help developers understand and analyze code changes.
For example, they can provide additional explanatory power in maintenance classification~\cite{HindleGGH09} and Just-In-Time defect prediction~\cite{BarnettGSM16}. Refactoring opportunities can be found with the analysis of commit messages~\cite{RebaiKASK20}.

However, writing commit messages manually is time-consuming and laborious especially when the code is updated frequently~\cite{Liu20Atom}. 
Generating commit messages automatically is very helpful to developers. 
Early work on commit message generation~\cite{VasquezCAP15,BuseW10,Cortes-CoyVAP14} is based on expert rules. Commit messages generated by rule-based methods tend to have too many lines, making it difficult to convey the key intention of the code changes~\cite{LiuXHLXW18}. 
Later, information retrieval techniques are introduced to commit message generation. 
For instance, \Nngen proposed by Liu et al.~\cite{LiuXHLXW18} is a simple yet effective retrieval-based method utilizing nearest neighbor algorithm.
\ChangeDoc proposed by Huang et al.~\cite{HuangJZCZT20} is another method that retrieves the most similar commits according to the syntax and semantics in the changed code. 
Recently, various deep learning-based models are  proposed for commit message generation. Some studies~\cite{JiangAM17,LoyolaMBMS18,LoyolaMM17} represent code changes as textural sequences and use Neural Machine Translation (NMT) techniques to translate the source code changes into target commit messages. 
In addition, Liu et al.~\cite{LiuLZFDQ19} adopt the pointer-generator network~\cite{SeeLM17} to handle the out-of-vocabulary (OOV) words. 
Other studies leverage the rich structural information of source code. Xu et al.~\cite{Xu00GT019} jointly model the semantic representation and structural representation of code changes, Liu et al.~\cite{Liu20Atom} capture both the AST structure of code changes and its semantics for commit message generation.

However, we notice that several important aspects are overlooked in existing work. 
Firstly, when evaluating commit message generation models, evaluation metrics being used vary a lot. The differences and applicability of these metrics have received little attention. 
Second, the evaluation datasets are different for different models and most studies experiment on only one dataset. 
Third, the applicable scenarios of the models are rarely discussed, such as data splitting, etc.
In view of the above limitations of existing work, in this paper, we would like to dive deep into the problem and answer: \emph{how to evaluate and compare commit message generation models more correctly and comprehensively?}

To answer the above question, we conduct a systematic analysis of commit message generation methods and their performance. 
We analyze the features in commit messages and compare the BLEU variants with human judgment. 
Moreover, we collect a large-scale commit message dataset from 500 repositories with more than 3.6M commit messages in five popular programming languages (PLs). 
Benefited from the comprehensive information provided by our dataset, we evaluate the performance of existing methods on multiple PLs. 
We also compare the influence of different splitting strategies on model performance and discuss the applicable scenarios.

Through extensive experimental and human evaluation, we obtain the following findings about the current commit message generation models and datasets: 

\begin{itemize}[topsep=0pt,itemsep=0pt,partopsep=0pt,parsep=0pt,leftmargin=8pt]

    \item Most existing datasets are crawled from only Java repositories while repositories in other PLs are rarely taken into account. Moreover, we find some context information contributing to generation models is not available in the existing public datasets.
    
    \item By comparing three BLEU variants commonly used in existing work, we find they show inconsistent results in many cases. From the correlation coefficient between human evaluation and different BLEU variants, we find that \bleumosesnorm is a more suitable BLEU variant for this task and it is recommended to be used in future research.
    
    \item By comparing different models on existing datasets and our multi-programming-language dataset, we find existing models show different performance. On the positive side, most of them can be migrated to repositories in other PLs.
    
    \item The dataset splitting strategies have a significant impact on the evaluation of commit message generation models. Many studies randomly split datasets by commit. Such a splitting strategy cannot simulate the Just-In-Time situation, where the training set contains no data later than that in the test set. Our results show that evaluating models on datasets split by timestamp shows much worse performance than split by commit. In the scenario of Just-In-Time, 
    
    we suggest splitting by timestamp for a practical evaluation of models. 
    Moreover, splitting data by project also leads to worse performance, meaning that the generalization ability of existing models for new repositories is limited. 
    
    Therefore, in order to evaluate models more comprehensively, it is suggested to also evaluate models on datasets split by project.

\end{itemize}

Through these findings, we give 
actionable suggestions on comprehensively evaluating commit message generation models. Then, we discuss future work from three aspects: metrics, information, different scenarios. Finally, we discuss some threats to validity. To summarize, the major contributions of this paper are as follows:

\begin{itemize}[topsep=0pt,itemsep=0pt,partopsep=0pt,parsep=0pt,leftmargin=8pt]

\item We perform a systematic evaluation of existing work on commit message generation and summarize our findings (including the inappropriate use of BLEU metrics and data splitting strategies, etc) which have not received enough attention in the past.

\item We develop a large dataset in multi-programming-language which contains comprehensive information for each commit and is publicly available\footnote{\url{https://doi.org/10.5281/zenodo.5025758}}.

\item We suggest ways for better evaluating 
commit message generation models. We also suggest possible research directions that address the limitations of existing models.

\end{itemize}

\section{Commit Message Generation}

\begin{figure}[t]
    \includegraphics[width=1.0\columnwidth]{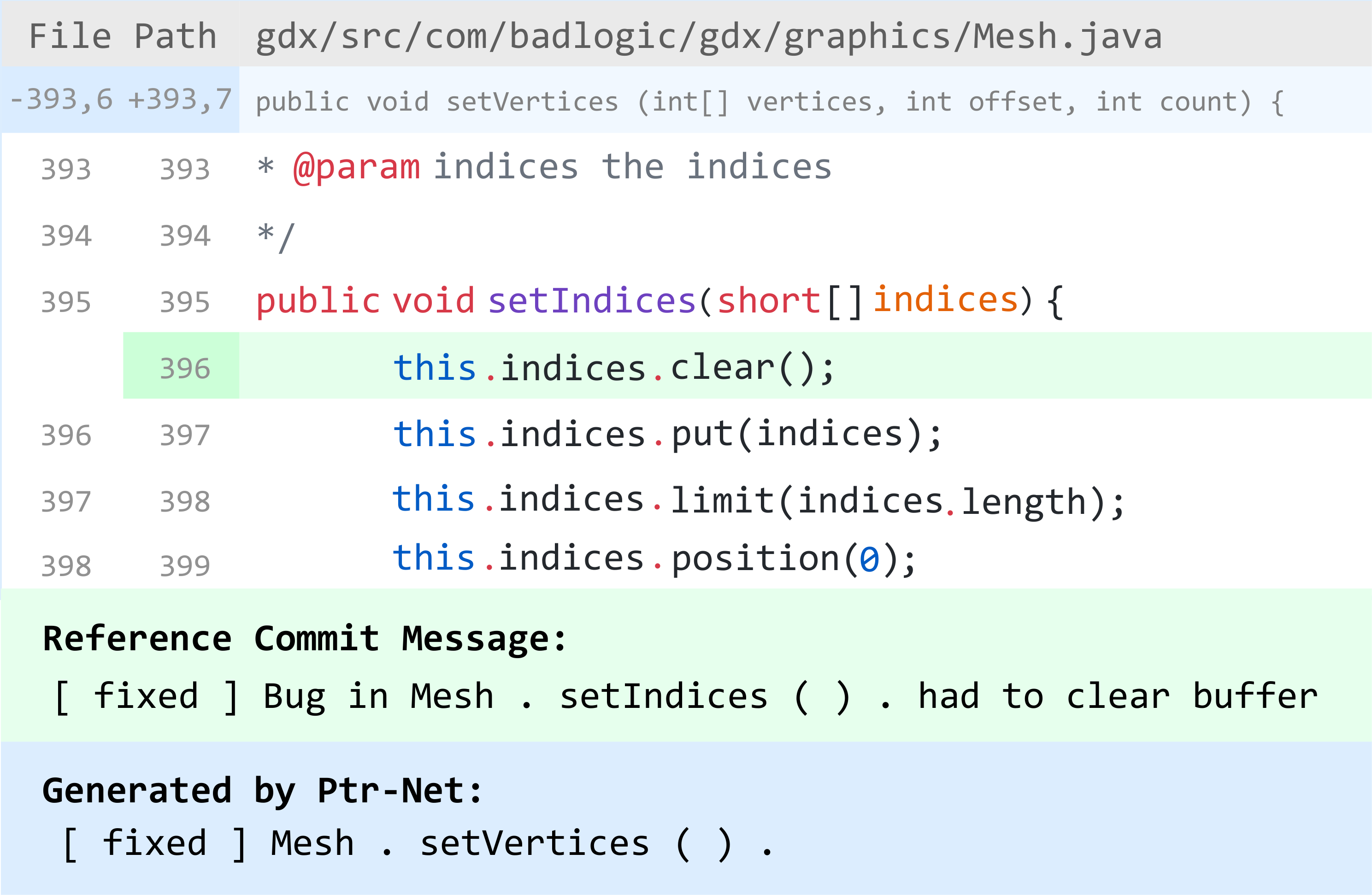}
    \vspace{-15pt}
    \caption{\textbf{Example of code diff and the corresponding commit message.}}
    
    \label{fig:example-code-diff}
    \vspace{-10pt}
\end{figure}

In this section, we give an overview of existing models, experimental datasets, and evaluation metrics for commit message generation.

\subsection{Models}

We study representative commit message generation models proposed in recent years. They can be classified into three categories, namely generation-based models, retrieval-based models, and hybrid models.

\subsubsection{Generation-based Models}
 
\begin{itemize}[topsep=0pt,itemsep=0pt,partopsep=0pt,parsep=0pt,leftmargin=8pt]

\item 
\textbf{\underline{\Commitgen}}, proposed by Jiang et al.~\cite{JiangAM17}, is an early attempt to adopt Neural Machine Translation (NMT) techniques in commit message generation. It treats code diffs and commit messages as inputs and outputs, respectively. \Commitgen adapts one NMT model Nematus~\cite{SennrichFCBHHJL17} with the attentional RNN encoder-decoder~\cite{BahdanauCB14}. The attention mechanism is introduced to capture long-distance features as \Commitgen uses 100 as the maximum input length, which is much longer than that (50) used in Nematus~\cite{SennrichFCBHHJL17}.

\item 
\textbf{\underline{\Nmt}}, proposed by Loyola et al.~\cite{LoyolaMM17}, is an encoder-decoder NMT model.
Different from \Commitgen which uses Bahdanau attention, it introduces the attention mechanism proposed by Luong et al~\cite{LuongPM15}. It is compared as a baseline model in ATOM~\cite{Liu20Atom}.

\item 
\textbf{\underline{\Codisum}}, proposed by Xu et al.~\cite{Xu00GT019}, is another encoder-decoder based model. The major difference between \Codisum and other models is the design of the encoder part. \Codisum jointly models the structure and the semantics of the code diffs with a multi-layer bidirectional GRU to better learn the representations of the code changes. Moreover, the copying mechanism~\cite{SeeLM17} is used in the decoder to mitigate the out-of-vocabulary (OOV) issue.

\item 
\textbf{\underline{\Ptrnet}}, proposed by Liu et al.~\cite{LiuLZFDQ19}, is an improved attentional RNN encoder-decoder model that incorporates the pointer-generator network~\cite{SeeLM17} to translate code diffs to commit messages. It learns to either copy an OOV word from the code or generate a word from the fixed vocabulary. 

\end{itemize}

\subsubsection{Retrieval-based Models}

\begin{itemize}[topsep=0pt,itemsep=0pt,partopsep=0pt,parsep=0pt,leftmargin=8pt]

\item \textbf{\underline{\Nngen}}, proposed by Liu et al.~\cite{LiuXHLXW18}, leverages the nearest neighbor (NN) algorithm to generate commit messages. Code diffs are represented as vectors in the form of ``bag of words''~\cite{Mogotsi10}. To generate a commit message, \Nngen calculates the cosine similarity between the target code diff and each code diff in the training set. Then, the top-k code diffs in the training set are selected to compute the BLEU scores between each of them and the target code diff. The one with the largest BLEU score is regarded as the most similar code diff and its commit message will be used as the target commit message. The use of cosine similarity for retrieval can boost efficiency and the use of BLEU for ranking can improve the performance. Therefore, \Nngen strikes a balance between effectiveness and efficiency.

\item \textbf{\underline{\CCVec}}, proposed by Hoang et al.~\cite{HoangK0L20}, is a neural model that learns a representation of code changes guided by the accompanying commit messages. It aims to represent the semantic intent of the code changes by modeling the hierarchical structure of a code change with an attention mechanism and using various comparison functions to identify the differences between the deleted and added code. The learned vector representations of diffs are used to adapt the \Nngen model to retrieve a code diff that is most similar to the input so that the corresponding commit message can be used as the output.

\end{itemize}

\subsubsection{Hybrid Model}

\begin{itemize}[topsep=0pt,itemsep=0pt,partopsep=0pt,parsep=0pt,leftmargin=8pt]

\item \textbf{\underline{\Atom}}, proposed by Liu et al.~\cite{Liu20Atom}, is a hybrid model that combines the techniques in generation-based models and retrieval-based models. \Atom is the first model that makes use of the Abstract Syntax Trees (ASTs) of the code diffs for commit message generation. In the generation module, AST paths extracted from ASTs are encoded with a BiLSTM model to represent the code diffs and then the attention mechanism is used in the decoder to generate a sequence as the commit message. In the retrieval module, the code diff that has the largest cosine similarity with the input code diff is retrieved from the training set. Finally, a hybrid ranking module is used to prioritize the commit messages obtained from the generation and retrieval modules.

\end{itemize}

\subsection{Datasets}\label{sec:bg:datasets}

There are several existing datasets for the commit message generation task. \Tab~\ref{tab:dataset} reports their basic information.

\begin{table}[t]

\footnotesize
\centering \setlength{\tabcolsep}{3.0pt}
\caption{Statistics of Datasets. Datasets evaluated are marked in bold.}
\vspace{\figmargin}
\begin{tabular}{cccp{0.3\columnwidth}<{\centering}} \toprule
Dataset                  & Lang.      & \# Commits     &  Content \\ \midrule 
\textbf{\CommitgenData}           & Java       & 32,208          & $\langle$Diff, Message$\rangle$  \\ \midrule
\textbf{\NngenData}               & Java       & 27,144          & $\langle$Diff, Message$\rangle$ \\ \midrule
\textbf{\CodisumData}             & Java       & 90,661          & $\langle$Diff, Message$\rangle$ \\ \midrule
\PtrgenData              & Java       & 32,663          & $\langle$Diff, Message$\rangle$ \\ \midrule
\multirow{4}{*}{\MlData} & Java       & 38,734          &   \\
                         \cmidrule(r){2-3}
                         & C++        & 43,868          & $\langle$Diff, Message,\\
                         \cmidrule(r){2-3}
                         & Python     & 41,023          & RepoName$\rangle$\\
                         \cmidrule(r){2-3}
                         & JavaScript & 29,821          & \\ \midrule
\multirow{3}{*}{\AtomData} & \multirow{3}{*}{Java}     & \multirow{3}{*}{197,968}         & $\langle$Diff, Message, \\
                         &          &                   & RepoName, SHA, \\
                         &          &                   & Timestamp$\rangle$ \\ \midrule
\textbf{\multirow{5}{*}{\mcmd (Ours)}} & Java  & 450,000         &   \\
                         \cmidrule(r){2-3}
                         & C\#        & 450,000         & $\langle$Diff, Message,\\
                         \cmidrule(r){2-3}
                         & C++        & 450,000         & RepoFullName, \\
                         \cmidrule(r){2-3}
                         & Python     & 450,000         & SHA,\\ 
                         \cmidrule(r){2-3}
                         & JavaScript & 450,000         & Timestamp$\rangle$\\ \bottomrule
\end{tabular}
\vspace{\figmargin}
\label{tab:dataset}

\end{table}

\begin{itemize}[topsep=0pt,itemsep=0pt,partopsep=0pt,parsep=0pt,leftmargin=8pt]

\item \textbf{\underline{\CommitgenData}}: It is an early commit message generation dataset used in \Commitgen and other studies~\cite{LiuXHLXW18,LiuLZFDQ19,HoangK0L20}. It is pre-processed from the commit dataset provided by Jiang et al.~\cite{JiangM17} which is collected from top-1000 Java GitHub projects. \CommitgenData extracts the first sentence from the original commit messages and removes the commits which have ids of issues. Merge and rollback commits are also removed because existing models are not suitable for most of these commits with too many lines. 
A Verb-Direct Object filter is also introduced to filter out non-compliant commit messages. 
After filtering, 32K commits remain in the dataset. The training, validation and test sets contain $26,208$, $3,000$, and $3,000$ commits, respectively.

\item \textbf{\underline{\NngenData}}: Liu et al.~\cite{LiuXHLXW18} find that \CommitgenData contains about 16\% noisy messages that can be divided into two categories: \textit{bot messages} (generated by bot) and \textit{trivial messages} (written by human but contain little information about code diff). 
Liu et al.~\cite{LiuXHLXW18} remove these noises and proposed the cleaned subset of \CommitgenData, i.e., \NngenData. The training, validation and test sets contain $22,112$ / $2,511$ / $2,521$  commits, respectively.

\item \textbf{\underline{\CodisumData}}: Based on the dataset in \cite{JiangM17}, Xu et al.~\cite{Xu00GT019} remove the commits that contain no source code changes. They also remove punctuations and special symbols in commit messages, and filter commit messages that contain less than three words and duplications. Finally, they obtain 90,661 pairs of $\langle$Diff, Message$\rangle$, and they randomly choose $75,000$ / $8,000$ / $7,661$ for training, validation, and testing.

\item \textbf{\underline{\PtrgenData}}: Liu et al.~\cite{LiuLZFDQ19} collect the top 1,001-2,081 
Java projects. They remove the rollback and merged commits, extract the first sentence from messages, and replace the diff signs ($+$ and $-$) with special tokens \texttt{<add>} and \texttt{<delete>}. 
The training, validation and test sets contain $23,623$ / $5,051$ / $3,989$ commits, respectively.

\item \textbf{\underline{\MlData}}: Loyola et al.~\cite{LoyolaMM17} collect pairs of $\langle$Diff, Message$\rangle$ from 12 public projects in 4 PLs: Python, JavaScript, C++, and Java, with 12,787 pairs of $\langle$Diff, Message$\rangle$ in each project on average. 
They randomly choose 90\% for training, 10\% for validation, and 10\% for testing.

\item \textbf{\underline{\AtomData}}: Liu et al.~\cite{Liu20Atom} collect data from 56 Java projects with the largest number of stars. After filtering commits with noisy messages and commits that contain no source code changes, \AtomData contains 197,968 commits. This dataset is designed to provide not only the raw commits but also the extracted functions which are affected in each commit. 
They randomly choose 81\% for training, 10\% for validation, and 9\% for testing.

\end{itemize}

\subsection{Evaluation Metrics}\label{sec:metrics}

Several metrics commonly used in NLP tasks such as machine translation, text summarization, and captioning can be adopted for evaluating commit message generation. These metrics include BLEU~\cite{PapineniRWZ02}, Meteor~\cite{BanerjeeL05}, Rouge-L~\cite{lin-2004-rouge}, Cider~\cite{VedantamZP15}, etc. In this study, we focus on BLEU as it is the metric that the related works~\cite{JiangAM17, LiuXHLXW18,  HoangK0L20,wang2020cocogum,GuZZK16,zhangretrieval20} use to evaluate the performance of commit message generation models. 

BLEU score is used to evaluate the correlation between the generated and the reference sentences in NLP tasks. For the commit message generation, the references are the commit messages written by developers and the generated sentences are the outputs from the models. Different BLEU variants are used in prior work and they could produce different scores for the same generated commit message. 
In the following, we are going to illustrate three different  variants of BLEU used before. 
The names are not intended to be a standard in the literature, but just for easy reference in this study. 

\begin{itemize}[topsep=0pt,itemsep=0pt,partopsep=0pt,parsep=0pt,leftmargin=8pt]

\item \textbf{\underline{\bleumoses}}: The evaluation described in \cite{JiangAM17, LiuXHLXW18, LiuLZFDQ19} use the same BLEU script from the open-sourced code in \cite{KoehnHBCFBCSMZDBCH07} which calculates \bleumoses. \bleumoses is designed for statistical translation, which does not use a smoothing function. It can be calculated as follows:

\vspace{-10pt}
\begin{equation}
\footnotesize
\mathrm{BLEU}=\mathrm{BP} \cdot \exp \left(\sum_{n=1}^{N} w_{n} \log p_{n}\right)
\label{equ:BLEU}
\end{equation}

where $w_n$ is the weight of \(n\)-gram precision $p_n$, which can be obtained as Equation~\ref{equ:precision}. If not explicitly specified, $N$ = 4 and uniform weights $w_n = 1/4$. 

$BP$ is brevity penalty which is computed as: 
\begin{equation}
\footnotesize
\mathrm{BP}=\left\{\begin{array}{ll}
1 & \text { if } c>r \\
e^{(1-r / c)} & \text { if } c \leq r
\end{array}\right.
\label{equ:BP}
\end{equation}
\noindent where \(c\) is the length of the candidate generation and \(r\) is the length of the reference.

The n-gram precision $p_n$ can be obtained as,
\begin{equation}
\footnotesize
\begin{aligned}
p_n={m_n}/{l_n} 
\end{aligned}
\label{equ:precision}
\end{equation}
\noindent where \(m_{n}\) is the number of matched n-grams between the reference and the generation, and \(l_{n}\) is the total number of n-grams in the generation.

\item \textbf{\underline{\bleumosesnorm}}: \bleumosesnorm is a BLEU variant adapted from \bleumoses. It is used by Loyola et al.~\cite{LoyolaMM17}. One  difference between \bleumosesnorm and \bleumoses is that \bleumosesnorm converts all  characters both in the reference and the generation to lowercase before calculating scores. Therefore, \bleumosesnorm is case insensitive. The smoothing method proposed by Lin and Och~\cite{DBLP:conf/acl/LinO04} is used in \bleumosesnorm to smooth the calculation of n-gram precision scores.  
It adds a constant number (one) to both the numerator and denominator of $p_n$ for $n > 1$.

\item \textbf{\underline{\bleunltksmfive}}: Hoang et al.~\cite{HoangK0L20} use BLEU measure provided by NLTK~\cite{Xue11} with the smoothing method proposed by Chen and Cherry~\cite{ChenC14} in their evaluation. 
This smoothing method~\cite{ChenC14} is inspired by the assumption that matched counts for similar values of \(n\) should be similar. The average value of the \(n-1\) , \(n\) and \(n+1\) –gram matched counts is used as  \(n\)-gram matched count. \(m_0'\) is defined as \(m_1+1\). Therefore, \(m_n'\) for \(n>0\) is defined as: $m_n'=(m_{n-1}'+m_n+m_{n+1})/{3}$.

\end{itemize}

\section{Study Design}

\subsection{Experimental Models}\label{sec:exp-models}

In this study, we select the models to be evaluated according to the following criteria: a) source code is publicly available, and b) we can confirm the correctness of source code by checking the implementation provided by authors and reproducing results presented in the original paper. 
\Commitgen~\cite{commitgenurl}, 
\Codisum~\cite{codisumurl}, 
\Nmt~\cite{nmturl}, 
\Ptrnet~\cite{ptrneturl}, 
and \Nngen~\cite{nngenurl} 
satisfy these criteria and thus are selected for in-depth evaluation in our study. For models that hyper-parameter settings are reported in their papers, we use the same hyper-parameters. Otherwise, we tune the hyper-parameters  empirically to optimize each model.

For {\CCVec}, after inspecting its public source code\footnote{\url{https://github.com/CC2Vec/CC2Vec}}, we suspect that the implementation is inconsistent with descriptions in the paper. We find that the scores reported in \CCVec paper are produced by the code that retrieves the commit message of the most similar code diff in terms of \emph{BLEU score} instead of the vector representation described in the \CCVec paper. We have tried to modify the code according to the paper, but the result drops significantly in BLEU. 
We have contacted the authors regarding this issue. 
Therefore, \CCVec is not evaluated in this study.

For {\Atom}, extracting AST paths for code diff is a key step during the pre-processing. However, the tool for this step is not available. We have contacted the authors but it cannot be provided for commercial reasons. 
We have tried to replace the required path extraction tool with JavaExtractor\footnote{\url{https://github.com/tech-srl/code2seq/tree/master/JavaExtractor}}~\cite{AlonBLY19} for extracting AST paths from the Java function. 
However, this attempt cannot fully match the results in \Atom paper, therefore, in most experiments except for the one reported in \Tab~\ref{tab:baselines1}, \Atom is not evaluated. 

\vspace{-5pt}
\subsection{Experimental Datasets}

\subsubsection{Existing Datasets}~\label{sec:existing_datasets} We select datasets from the existing ones based on their availability and representativeness. 
In this way, three existing datasets are chosen as highlighted in bold in \Tab~\ref{tab:dataset}: {\CommitgenData}, {\NngenData}, and {\CodisumData}. The reasons are explained as follows.

\CommitgenData and \NngenData are Java datasets and commonly used in ~\cite{JiangM17,LiuXHLXW18,LiuLZFDQ19,HoangK0L20}. 
\NngenData is more difficult than \CommitgenData because commit messages with certain patterns are filtered in \NngenData. 
\CodisumData is another Java dataset with different features compared to  \CommitgenData and \NngenData. It is a deduplicated dataset, i.e., its training set and test set are not overlapping. \CodisumData can be used to evaluate the generalization ability of models.

\PtrgenData is not used in our study since it is very similar to \CommitgenData except that \CommitgenData is collected from top-1,000 starred GitHub projects while \PtrgenData is from top-1001 to top-2081 projects. The performance difference is reported to be very small on \CommitgenData and  \PtrgenData~\cite{LiuLZFDQ19}. Besides, \CommitgenData is used more often in the literature~\cite{JiangAM17,LiuXHLXW18,Liu0T0L19,HoangK0L20} than \PtrgenData~\cite{Liu0T0L19}.

\MlData is not used for two reasons. 
First, in \MlData, commits for each programming language are collected from only three repositories, resulting in small and sparse data. Second, the three collected repositories are not available from the provided link\footnote{\url{https://osf.io/67kyc/?view_only=ad588fe5d1a14dd795553fb4951b5bf9}}, making it difficult to inspect the data source.

\AtomData is not chosen because the given dataset is incomplete. For example, all commits to the repository \texttt{retrofit} are missing. Recovering the data is not feasible because both the repositories' full names (including owner and repositories' name) and version numbers are not provided.

\subsubsection{Our Dataset \mcmd}
Existing datasets have facilitated the development of commit message generation. However, the available datasets have their limitations: most are in a single programming language (i.e., Java), and the available information is very limited.
There is only one dataset \MlData~\cite{LoyolaMM17} in multiple PLs, but it is not usable as explained in \Sec~\ref{sec:existing_datasets}. 
To provide a large-scale dataset in multiple PLs and with rich information, we created a new dataset \textbf{\mcmd}, short for \textbf{M}ulti-programming-language \textbf{C}ommit \textbf{M}essage \textbf{D}ataset. For each language, we collected commits before 2021 from the top 100 starred projects on GitHub. In this step, a total of 3.69M commits were collected. 
We removed branch merging and rollback commits, and filtered out noisy messages as Liu et al.~\cite{LiuXHLXW18}, to improve the quality of commits in our dataset. 
About 3.42M commits remain after filtering. To balance the size of data in each programming language so that we can fairly compare the performance of models in different programming language in subsequent experiments, we randomly sampled and retained 450,000 commits for each language. 

Existing datasets~\cite{JiangAM17,LiuXHLXW18,Xu00GT019} contain the information of only code diffs and the commit messages. 
However, the context of the code diffs can contribute to explaining why this code is added and what role it plays in the software~\cite{SillitoMV08}. 
For example, extracting AST paths from the code diffs is beneficial to the commit message generation model~\cite{Liu20Atom}, which requires the dataset to provide enough information to find the complete affected functions around the code changes. However, most of the existing datasets~\cite{JiangAM17,LiuXHLXW18,Xu00GT019,LoyolaMM17} do not provide information for retrieving related functions. To trace back to the original repository, the \emph{RepoFullname} (including owner and repository's name) and \emph{SHA} of a repository should be recorded. The RepoFullname can be used to find the corresponding repository and SHA is a unique ID to identify the version of the repository. 
Moreover, if we want to split the dataset by timestamp, timestamps of commits are necessary. Considering the above demands, our dataset \mcmd contains the complete information of commits, including not only code diffs and commit messages, but also RepoFullname, SHA, and timestamp. We have made \mcmd public\footnote{\url{https://doi.org/10.5281/zenodo.5025758}}  to benefit future research on commit message generation.

\vspace{-5pt}
\subsection{Research Questions}

We have identified the following Research Questions (RQs) and will seek their answers in our evaluation:

\vspace{5pt}

\noindent \textbf{RQ1: How do different BLEU variants affect the evaluation of commit message generation?}

As described in \Sec~\ref{sec:metrics}, most existing works use BLEU as an evaluation metric but the scores in their papers are different BLEU variants. 
Scores for different BLEU variants can vary a lot for the same sentence as shown in \Tab~\ref{tab:metrics_compare}. 

\begin{table*}

\centering

\caption{Example scores under different metrics.}

\begin{tabular}{ccccc} \toprule
\textbf{References} & \textbf{Generated} & \textbf{\bleumoses}  & \textbf{\bleumosesnorm} & \textbf{\bleunltksmfive} \\ \midrule
 add setup ( ) & add setUp ( ) & 0.00 & 100.00 & 22.80 \\  \midrule
Fix merge conflicts & fix merge conflicts & 0.00 & 100.00 & 25.00 \\  \midrule
BAEL - 3001 & BAEL - 2412 : Add a new class & 0.00 & 19.64 & 12.54 \\  \midrule
Fix typo & Fix typo in core - validation . adoc & 0.00 & 19.64 & 12.54 \\ \midrule
 Update visualvm to build 908 & [ GR - 6405 ] Update visualvm . & 0.00 & 19.64 & 12.54 \\ \midrule
\multirow{2}{*}{[ FIXED JENKINS - 12514 ]} & \multirow{2}{*}{\shortstack{[ FIXED JENKINS - 12514 ] Fixed a bug \\ in bundled plugins on Windows .}} & \multirow{2}{*}{22.31} & \multirow{2}{*}{36.41} & \multirow{2}{*}{41.26} \\ \\ \midrule

\multirow{1}{*}{\shortstack{[ GR - 22084 ] Add TruffleCreateGraphTime timer .}} & \multirow{1}{*}{Add a timer to the timer .} & \multirow{1}{*}{0.00} & \multirow{1}{*}{19.68} & \multirow{1}{*}{11.51} \\  \midrule
 Remove dead code .  & [ GR - 19154 ] Remove unused code .   & 0.00 & 19.07 & 13.25  \\   \midrule
Fix reported leaks  & Fix a bug in SnappyFramedEncoderTest  & 0.00 & 24.03 & 8.98   \\  \midrule
{[ fixed ] Bug in Mesh . setIndices ( ) . had to clear buffer first .} & {[ fixed ] Mesh . setVertices ( ) .} & 0.00 & 18.97 & 16.63 \\
\bottomrule
\end{tabular}

\label{tab:metrics_compare}
\vspace{-5pt}
\end{table*}

For instance, as \Tab~\ref{tab:metrics_compare} and \Fig~\ref{fig:example-code-diff} illustrate, the commit message ``\texttt{[ fixed ] Mesh . setVertices ( ) .}'' generated by \Ptrnet is relatively reasonable for that code changes. Compared with the reference, it has shared tokens and the meaning is partially correct. However, it has different scores for different BLEU variants, as shown in \Tab~\ref{tab:metrics_compare}. The \bleumoses score of 0 means that this generation and reference are completely different, which is not true. Another case is the commit message ``\texttt{fix merge conflicts}'' generated by \Nmt, which has the same meaning with the reference but it has lower scores for \bleumoses (0) and \bleunltksmfive (25.00) while it has the perfect score for \bleumosesnorm (100). These examples are just the ``tip of the iceber’’.

RQ1 chooses the most suitable BLEU metric for the task of commit message generation by human evaluation and analyzes why that variant is better than others.
Following best practice for the human evaluation~\cite{LeeGMWK19}, three human experts manually labeled the data. All of them have more than 5 years of programming experience and they are majored in Computer Science.
Firstly, we select 100 commit messages randomly from generation results which show large disagreement (the variance among the three BLEU metrics is larger than 30). Then, we define five levels of criteria for manual labeling as shown in \Tab~\ref{tab:human_score_meaning}. The raters give a score between 0 to 4 to measure the semantic similarities between reference and the generated commit message. 
After labeling, all scores are double checked by each rater to confirm whether scores from human are stable and reliable. 

To validate the reliability of human scores, we calculate Krippendorff's alpha~\cite{hayes2007answering} and Kendall rank correlation coefficient (Kendall's Tau)~\cite{kendall1945treatment} values\footnote{The details of calculation can be seen at our repository~\url{https://github.com/DeepSoftwareAnalytics/CommitMsgEmpirical}}.
Before the calculation, these human direct assessments are converted into relative rankings as Direct Assessment Relative Rankings (DaRR) serve as the golden standard for segment-level evaluation~\cite{MaWBG19}.
The Krippendorff's alpha of the three raters is 0.86, and the Kendall's Tau value between any two raters is greater than 0.8, which indicates there is a high degree of agreement between the raters and human scores are reliable.

Human scores are regarded as the reference and we want to compare the three BLEU variants by their correlations with the reference.
Spearson~\cite{zwillinger1999crc} and Kendall~\cite{kendall1945treatment} are selected because the scores are ordinal and satisfied their assumptions.  
Note that these correlation coefficients are calculated per commit
and more details can be found in our repository.

\begin{table}[t]
\centering 

\footnotesize
\caption{The meaning of scores in human evaluation.}
\vspace{\figmargin}
\begin{tabular}{l|l} \toprule
{Score}   & {Meaning}  \\ \midrule 
0       & No similarity between the generation and reference. \\
1       & Have few shared tokens, not semantically similar. \\
2       & Have some shared tokens, probable semantically similar. \\
3       & Much similar in semantic but a few tokens are different. \\
4       & Identical in semantic. \\ \bottomrule
\end{tabular}

\vspace{-10pt}
\label{tab:human_score_meaning}
\end{table}

\noindent \textbf{RQ2: How good are the existing models and datasets?}

As described in \Sec~\ref{sec:bg:datasets}, When evaluating commit message generation models, not only are the evaluation metrics different, but the datasets used are also different. Most of the existing studies~\cite{LoyolaMM17,Xu00GT019,Liu20Atom} only experiment on one dataset. Therefore, in RQ2, we conduct a unified evaluation of existing models on existing datasets, and study the impact of using different datasets on model evaluation.

\noindent \textbf{RQ3: Why do we need a new dataset \mcmd for evaluating commit message generation?}\label{sec:rq3}

Since most of the prior works focus on Java datasets, there is a lack of research on other PLs' repositories that have the same need to generate commit messages automatically.  In RQ3, we explore the necessity of a new dataset, and rely on the new dataset \mcmd we collected to explore the performance of migrating existing models to other PLs. Similar to previous work~\cite{JiangAM17,Liu0T0L19,Liu20Atom}, for the dataset of each language in \mcmd, we randomly select 80\% data for training, 10\% data for validation and 10\% data for testing.

\noindent \textbf{RQ4: What is the impact of different dataset splitting strategies?}\label{sec:rq4}

Commit message generation models have different usage scenarios, so they need to be evaluated on datasets split by different strategies that simulate different scenarios. However, for the datasets used in most previous work, commits are split into training, validation, and testing sets randomly (split by commit),
while other situations such as split by project (where training, validation, and testing sets contain commits from disjoint projects) are not considered. 
In RQ4, we study the impact of different dataset splitting strategies from two aspects.

\textbf{Split by timestamp.} As a previous study on code summarization~\cite{LeClairM19} suggests: ``Care must be taken to avoid unrealistic scenarios, such as ensuring that the training set consists only of code older than the code in the test set''. The commit message generation task is similar to code summarization task in this regard, especially in a just-in-time (JIT) scenario, the model cannot see future data and can only use past data for training. 

Therefore, we further conduct experiments on datasets split by timestamp instead of by commit to ensure future commits are not used as training data. 
In this splitting strategy, we divide each programming language's dataset for training (80\%), validation (10\%), and test (10\%) in chronological order.

\textbf{Split by project.} 
According to the study of LeClair and McMillan~\cite{LeClairM19} on code summarization, splitting dataset by function (in analogy with “commit” in our study) might cause information leakage from test set projects into the training or validation sets and should be avoided. Following previous studies~\cite{LeClairM19, Liu20Atom}, we also evaluate the performance of models based on \mcmd  split by project to ensure that projects in the training, validation, and test sets are disjoint. 
In this splitting strategy, we divide each PL's dataset for training (80\%), validation (10\%), and test (10\%) by the repository.

\begin{figure}[t]
    \centering \normalsize
    \includegraphics[width=0.8\columnwidth]{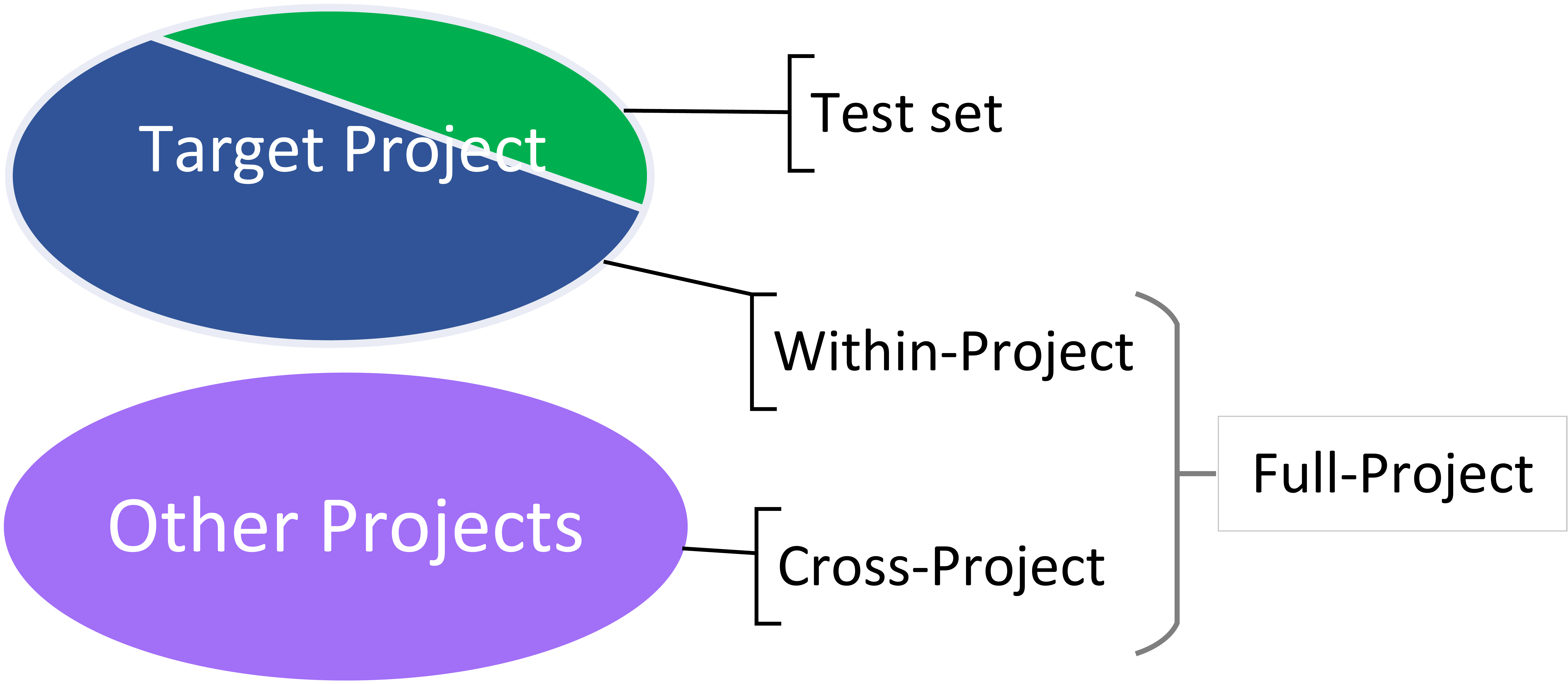} 
    \vspace{\figmargin}
    \caption{Illustration of data settings for single project experiments in \Tab~\ref{tab:single-project}. The training data for Within-Project, Cross-Project, and Full-Project settings are blue, purple, and blue+purple parts, respectively. Test data (the green part) is the same for all three settings, taken from the target project.} 
    \vspace{\figmargin}
    \label{fig:single-projct}

\end{figure}

The experiments of splitting by project can reflect the performance of models on new repositories.
For a repository that has commits before, the models can be trained by using only other repositories, using the repository itself, or using both of itself and others. 
Furthermore, to mimic the scenario that we are predicting commit messages on a new project with a trained model, we conduct a series of experiments called \emph{single project experiments}. As illustrated in \Fig~\ref{fig:single-projct}, the test set for all experiments 
is the same, and it comes from the target project. The training sets for the three experiments in the single project experiments are: \textcircled{1} data only from the target project for \emph{Within-Project} setting; \textcircled{2} data only from other projects in the same programming language for \emph{Cross-Project} setting; and \textcircled{3} the union of training sets from the target project and other projects for \emph{Full-Project} setting.
The experiment on these three settings can provide suggestions for the models' usability on existing repositories.
\vspace{-5pt}

\section{Results and Findings}\label{sec:results}

\subsection{How Do Different BLEU Variants Affect the Evaluation of Commit Message Generation? (\textbf{RQ1})}\label{sec:metrics_result}

To answer this question, we evaluate commit message generation models using different BLEU variants, and compare BLEU scores with the results from a human evaluation.

\subsubsection{Experiments under different BLEU variants}

\begin{table*}[t]
\centering \footnotesize
\caption{Models performance. BM, BN, BC are short for \bleumoses, \bleumosesnorm, and \bleunltksmfive, respectively.} 
\vspace{\figmargin}
\begin{tabular}{>{\bfseries}l *{3}{l} *{3}{l} *{3}{l} *{3}{l}} \toprule

\multirow{2}{*}{Model} & \multicolumn{3}{c}{\CommitgenData } & \multicolumn{3}{c}{\NngenData } & \multicolumn{3}{c}{\CodisumData  } & \multicolumn{3}{c}{\mcmdjava} \\
\cmidrule(r){2-4} \cmidrule(r){5-7}  \cmidrule(r){8-10} \cmidrule(r){11-13} 
& BM & BN & BC& BM & BN & BC & BM & BN & BC & BM & BN & BC \\
\midrule
\Commitgen   & 34.03&31.11&26.00  & 15.08&21.60&15.48   & 1.33 & 9.37 & 4.17  & 6.29& 12.39& 7.67 \\
\Codisum     & 0.00 & 6.88 & 0.49   & 0.00 & 8.03 &0.86  & 1.74 & 15.45 & 5.72    &2.00& 14.00& 5.37  \\
\Nmt         & 32.09&26.66&21.51  & 7.46&13.82&8.28     & 1.32 & 9.93 &3.81   & 9.17&13.39&10.24 \\
\Ptrnet      & 35.41&29.86&24.82  & 9.69&18.96&10.39    & 0.81&12.71&4.77 & 8.25& 15.33& 11.70\\
\Nngen       & 38.54&34.74&29.44  & 16.41&23.07&16.77   & 3.04 & 9.07 & 5.27  & 13.30& 17.81& 14.46\\

\Atom        & \multicolumn{1}{c}{/} & \multicolumn{1}{c}{/} & \multicolumn{1}{c}{/}              & \multicolumn{1}{c}{/} &\multicolumn{1}{c}{/} & \multicolumn{1}{c}{/}         &\multicolumn{1}{c}{/} &\multicolumn{1}{c}{/} &  \multicolumn{1}{c}{/}        & 7.47 & 16.42 & 9.29\\

\bottomrule
\end{tabular}
\vspace{\figmargin}

\label{tab:baselines1}

\end{table*}

As shown in~\Tab~\ref{tab:baselines1}, rankings of models are inconsistent under different metrics\footnote{All the generated commit messages are available in our repository.}. For example, on \CodisumData, \Codisum is the best model when measured by \bleumosesnorm or \bleunltksmfive, while \Nngen is the best model when measured by \bleumoses. On \CommitgenData, if the \bleumoses metric is used, \Ptrnet is better than \Commitgen, but an opposite conclusion will be drawn if \bleumosesnorm is used. Similar inconsistencies can also be observed in \Tab~\ref{tab:baselines-multilan}.

\subsubsection{Human Evaluation}

\Tab~\ref{tab:human-evaluation} shows the correlation scores between three BLEU variants and human evaluation, under two correlation metrics: 
Spearson~\cite{zwillinger1999crc} and Kendall~\cite{kendall1945treatment}. 
We can see that \bleumosesnorm is most correlated with human judgement and the conclusion consistently holds for all correlation metrics at the confidence level 95\%. 
After manually investigating a great number of commit messages in the test set and comparing the design of three BLEU variants, we find two possible reasons: smoothing and case sensitivity.

\begin{table}[t]
\centering \footnotesize
\caption{Correlation between metrics and human evaluations \\(with p-values shown in parentheses).}
\vspace{\figmargin}

\begin{tabular}{c|ccc} \toprule
Metric                & \bleumoses        & \bleumosesnorm                    & \bleunltksmfive       \\ 
\midrule
Spearman        & 0.1989 (\(<\) 0.05)    & \textbf{0.6188} (\(<\) 0.05)     & 0.5375 (\(<\) 0.05)  \\ 
Kendall         & 0.1716 (\(<\) 0.05)    & \textbf{0.4639} (\(<\) 0.05)     & 0.3967 (\(<\) 0.05)  \\ 

\bottomrule
\end{tabular}
\vspace{\figmargin}
\label{tab:human-evaluation}

\end{table}

\bleumoses does not perform smoothing when calculating each $p_n$ while \bleumosesnorm and \bleunltksmfive do so. 
However, more than 17.19\% commit messages have less than five tokens in \mcmd. Therefore, 4-gram precision $p_4$ (shown in Equation~\ref{equ:precision}) of these commit messages is close to zero, leading the geometric mean of n-gram precision scores to be zero even if there are many 1-gram, 2-gram, or 3-gram matches. 
Without smoothing, a short commit message that is identical to the reference will get a near-zero \bleumoses score, which is unreasonable. As many commit messages are short, we believe that  \bleumoses is not very suitable for evaluating commit message generation.

\bleumosesnorm is not case sensitive while \bleumoses and \bleunltksmfive are. As shown in~\Tab~\ref{tab:metrics_compare}, the generated message ``fix merge conflicts'' has the same meaning as the reference ``Fix merge conflicts''. The only difference is the case of ``Fix'' and ``fix''. Besides, other words such as ``add'' and ``Add'' also have the same meaning. Scores of \bleumoses and \bleunltksmfive for commit messages that differ only in case tend to be low. 
The low scores are unreasonable, since these messages have exactly the same meaning as the references.

\begin{tcolorbox}
\textbf{Summary:} 
For the evaluation of commit message generation models, using different metrics may lead to different conclusions. \bleumosesnorm, which uses a smoothing method and is case insensitive, is more in line with human judgments.
\end{tcolorbox}

\subsection{How Good Are the Existing Models and Datasets? (\textbf{RQ2})}\label{sec:RQ2}

Based on the experimental results shown in \Tab~\ref{tab:baselines1}, we have the following findings:
\begin{itemize}[topsep=0pt,itemsep=0pt,partopsep=0pt,parsep=0pt,leftmargin=8pt]

\item The scores of the same model on different datasets can vary a lot. For example, the \bleumosesnorm score of the \Nngen model on \CommitgenData is 34.74. When evaluated on \CodisumData, the score drops to 9.07. 
Hence, we should consider more datasets in the evaluation: good performance on one dataset does not mean we can observe similar performance on another dataset.

\item The scores on \CommitgenData are higher than scores on \NngenData. \NngenData is a subset of \CommitgenData, in which noisy messages are filtered out as described in \Sec~\ref{sec:bg:datasets}.
To investigate the role of noise data in model training and evaluation, 
we conduct ablation experiments on \Commitgen and the results are shown in \Tab~\ref{tab:different_testset}. We split the test set of \CommitgenData into 2 parts: one only contains noisy messages, and the other is the rest (i.e., the test set of \NngenData). 
We can see that the scores on the test set of noisy messages are much higher than that of \NngenData, indicating that noisy messages are easy to generate. 
However, these messages (e.g., branch merging messages) are often bot generated and do not need to be predicted by a model. 
Therefore, what really needs to be compared is the performance on the \NngenData test set. 
We further investigated whether excluding noisy data in model training will improve its performance. As shown in \Tab~\ref{tab:different_testset}, training on \NngenData (noise-free data) has a higher score than \CommitgenData, which indicates that it is better not to include noisy data in model training.

\item When experimenting on \CommitgenData and \NngenData, \Nngen has the highest scores under all metrics. But \Nngen does not perform the best on \CodisumData.  We speculate that this is because \Nngen is retrieval-based and it is easy for it to achieve high score on datasets with duplicated data. 
After checking, we find that in the test set of \NngenData, there are 16.02\% duplicated commit messages and 5.16\% duplicated $\langle$Diff, Message$\rangle$ pairs from the training set. And in the test set of \CommitgenData, there are 29.13\% duplicated commit messages and 4.67\% of duplicated $\langle$Diff, Message$\rangle$ pairs. 
In contrast, \CodisumData is a deduplicated dataset as described in \Sec~\ref{sec:existing_datasets}.
As a retrieval-based model, \Nngen obtains a high score by leveraging the duplication in the dataset.

\end{itemize}

\begin{table}[t]
\centering \footnotesize \setlength{\tabcolsep}{3.0pt}
\caption{Noisy data ablation study. BM, BN, BC are short for \bleumoses, \bleumosesnorm, and \bleunltksmfive. Model evaluated is \Commitgen.} 
    \vspace{\figmargin}
\begin{tabular}{>{\bfseries}l *{3}{l} *{3}{l}} \toprule
\multirow{2}{*}{\diagbox{Training}{Testing}}     & \multicolumn{3}{c}{\NngenData } & \multicolumn{3}{c}{Noisy data} \\ 
\cmidrule(r){2-4} \cmidrule(r){5-7}  
& BM & BN & BC & BM & BN & BC \\
\midrule
\CommitgenData  & 11.38 & 19.09 & 12.86   & 97.44 & 94.43 & 95.00   \\
\NngenData      & 15.08 & 21.60 & 15.48   & \multicolumn{1}{c}{/}     & \multicolumn{1}{c}{/}     & \multicolumn{1}{c}{/}       \\
\bottomrule
\end{tabular}
    \vspace{\figmargin}
\label{tab:different_testset}

\end{table}

\begin{tcolorbox}
\textbf{Summary:} 
More datasets can be used for more comprehensive evaluation since good performance of a model on one dataset does not mean good performance on other datasets. Removing noisy data (commits with bot and trivial messages) during model training can improve performance. 
The duplication of commit data makes the performance of retrieval-based models such as \Nngen better.

\end{tcolorbox}

\subsection{Why Do We Need a New Dataset \mcmd for Evaluating Commit Message Generation? (\textbf{RQ3}).}\label{sec:dataset}

From the results shown in \Tab~\ref{tab:baselines-multilan} and \Tab~\ref{tab:baselines1}, we have the following findings:

\begin{table*}[t]
\centering \footnotesize \setlength{\tabcolsep}{2.8pt}
\caption{Model performance on our dataset \mcmd. BM, BN, BC are short for \bleumoses, \bleumosesnorm, and \bleunltksmfive, respectively.}
\vspace{\figmargin}
\begin{tabular}{>{\bfseries}l *{3}{l} *{3}{l} *{3}{l} *{3}{l} *{3}{l} |*{3}{l}} \toprule
\multirow{2}{*}{Model}    & \multicolumn{3}{c}{\mcmdjava}  & \multicolumn{3}{c}{\mcmdcsharp}           & \multicolumn{3}{c}{\mcmdcpp}
                & \multicolumn{3}{c}{\mcmdpython}& \multicolumn{3}{c}{\mcmdjs}    & \multicolumn{3}{c}{Avg.} \\ 
                
    \cmidrule(r){2-4} \cmidrule(r){5-7}  \cmidrule(r){8-10} \cmidrule(r){11-13} \cmidrule(r){14-16} \cmidrule(r){17-19} 
    & BM & BN & BC & BM & BN & BC & BM & BN & BC & BM & BN & BC & BM & BN & BC & BM & BN & BC \\  
\midrule
\Commitgen      & 6.29& 12.39& 7.67         & 16.87& 18.14& 14.84               & 7.08& 11.58& 8.83
                & 5.81& 11.10& 7.78         & 11.66& 17.40& 11.81               & 9.54& 14.12& 10.19
                \\
\Codisum        & 2.00& 14.00& 5.37         & 1.78& 12.73& 4.74                 & 3.99& 12.46& 5.71
                & 2.63& 14.63& 5.74         & 1.51& 11.24& 4.31                 & 2.38& 13.01& 5.18
                \\
\Nmt            & 9.17& 13.39& 10.24        & 20.18& 17.32& 16.35               & 8.37& 11.56& 9.63
                & 7.64& 11.53& 9.56         & 15.00& 17.08& 14.77               & 12.07& 14.18& 12.11
                \\
\Ptrnet         & 8.25& 15.33& 11.70        & 18.92& 19.72& 17.18               & 5.91& 13.07& 9.97
                & 8.62& 15.99& 11.87        & \textbf{15.15}& \textbf{19.58}& \textbf{16.50}               & 11.37& 16.74& 13.44
                \\ 

\midrule

\Nngen          &\textbf{13.30} &\textbf{17.81} &\textbf{14.46}       
                &\textbf{22.79} &\textbf{22.92} &\textbf{20.73}               
                &\textbf{9.57}  &\textbf{13.69} &\textbf{10.89}
                &\textbf{11.66} &\textbf{16.64} &\textbf{13.19}       
                & 14.05& 18.03& 14.72               
                & \textbf{14.27}& \textbf{17.82}& \textbf{14.80}
                \\

\bottomrule
\end{tabular}
\vspace{\figmargin}
\label{tab:baselines-multilan}
\vspace{-5pt}
\end{table*}

\begin{itemize}[topsep=0pt,itemsep=0pt,partopsep=0pt,parsep=0pt,leftmargin=8pt]

    \item Most of existing datasets only retain $\langle$Diff, Message$\rangle$ information and cannot be used to evaluate models that require more information. For example, as shown in \Tab~\ref{tab:baselines1}, only our \mcmd dataset can be used to evaluate \Atom. This is because \Atom needs to know the complete code of the modified functions in order to extract the AST information. However, this information is unavailable in existing public datasets\footnote{Although the \AtomData can be used for evaluating ATOM, it is not publicly available as described in \Sec~\ref{sec:existing_datasets}.}. 
    Compared to existing datasets, our dataset \mcmd provides complete information for each commit. For example, the provided $\langle$RepoFullname, SHA$\rangle$ information can be used to obtain the complete functions for AST extraction.
    Please note that the \Atom results we presented here are only to illustrate the necessity of a richly informative dataset. As part of the \Atom code is not disclosed (as described in \Sec~\ref{sec:exp-models}), \Atom is out of the scope of this study. 
    We believe future research that requires the use of other commit information can benefit from the complete information provided by \mcmd.

    \item Extremely low scores are observed for \Codisum on \CommitgenData and \NngenData. The reason might be that the size of \CommitgenData and \NngenData is not large enough to support the model's training after filtering the data. \Codisum is designed to extract additional structure information from code changes in ``.java'' files. Therefore, code changes that are not related to ``.java'' files are filtered. After filtering by \Codisum, there are only hundreds of commits left, which are inadequate for its training. With the  large-scale dataset  \mcmd, we have tens of thousands of commits after filtering to support the training of \Codisum. Considering that some filtering steps reduce the size of the original dataset for the model's training, using a larger dataset can reduce the negative impact of insufficient training during the evaluation.

    \item The multiple-programming-language nature of the \mcmd dataset makes it possible to study the migration of commit message generation models to other PLs. From \Tab~\ref{tab:baselines-multilan}, we find that the ranking of models on the Java dataset cannot be preserved when migrating to other languages and the best model for different languages may vary.
    Overall, the retrieval-based model \Nngen performs the best, with an average \bleumosesnorm score of 17.82.
    But no model can consistently outperform others. 
    
\end{itemize}

\begin{tcolorbox}
\textbf{Summary:} 
A large-scale, multi-language, 
and information-rich dataset is needed to comprehensively evaluate commit message generation models. 
Overall, \Nngen performs the best, 
but no model can consistently outperform other models in all PLs. 
Therefore, when choosing commit message generation models for a new language, we suggest testing multiple models in that language to select the best one. 

\end{tcolorbox}

\subsection{What Is the Impact of Different Dataset Splitting Strategies? (\textbf{RQ4}).}\label{sec:splitting}

We analyze the impact of different dataset splitting strategies including splitting by timestamp and splitting by project.

\subsubsection{Split by Timestamp}

\Tab~\ref{tab:baselines_time} shows experimental results on datasets split by timestamp. Compared to \Tab~\ref{tab:baselines-multilan}, the performance of all models on all datasets drop consistently, and the BLEU scores of all models drop by 17.88\% - 51.71\% on average. 
This shows that it is more difficult to predict future commit messages
based on past data training. 

Although the retrieval-based model \Nngen shows the best performance in the split by commit setting as shown in \Tab~\ref{tab:baselines-multilan}, the results of splitting by timestamp are different. The performance degradation of \Nngen is greater than other models, and \Ptrnet performs the best. Therefore, in the JIT application scenario, it is not suitable to use datasets split by commit to evaluate models. Moreover, \Ptrnet which is based on generation and pointer-generator mechanism has better generalization ability in the JIT scenario.

\begin{table*}[t]
\centering \footnotesize \setlength{\tabcolsep}{5pt}

\caption{The performance on \mcmd split by timestamp under \bleumosesnorm. The score drop rates compared to \Tab~\ref{tab:baselines-multilan} are shown in parentheses.}
\vspace{\figmargin}
\begin{tabular}{>{\bfseries} l *{5}{l} | l} \toprule
Model   & \mcmdjava                         & \mcmdcsharp                             & \mcmdcpp                           
        & \mcmdpython                       & \mcmdjs                        & Avg. \\ 
        \midrule
\Commitgen & 8.08 (\textcolor{black}{$\downarrow$34.75\%})     & 4.53 (\textcolor{black}{$\downarrow$75.04\%})        & 7.08 (\textcolor{black}{$\downarrow$38.85\%})
        & 5.50 (\textcolor{black}{$\downarrow$50.49\%})         & 8.91 (\textcolor{black}{$\downarrow$48.77\%})        & 6.82 (\textcolor{black}{$\downarrow$51.71\%})
        \\
\Codisum & 12.71 (\textcolor{black}{$\downarrow$9.21\%})       & 4.85 (\textcolor{black}{$\downarrow$61.90\%})        & 12.24 (\textcolor{black}{$\downarrow$1.77\%})
        & 12.46 (\textcolor{black}{$\downarrow$14.83\%})       & 11.17 (\textcolor{black}{$\downarrow$0.62\%})        & 11.17 (\textcolor{black}{$\downarrow$17.88\%})
        \\
\Nmt    & 9.50 (\textcolor{black}{$\downarrow$29.05\%})        & 5.15  (\textcolor{black}{$\downarrow$70.27\%})        & 8.53 (\textcolor{black}{$\downarrow$26.21\%})    
        & 7.31  (\textcolor{black}{$\downarrow$36.60\%})        & 11.58 (\textcolor{black}{$\downarrow$32.20\%})        & 8.41 (\textcolor{black}{$\downarrow$40.65\%}) 
        \\
\Nngen  & 10.73 (\textcolor{black}{$\downarrow$39.78\%})        & 7.83 (\textcolor{black}{$\downarrow$65.81\%})         & 9.30 (\textcolor{black}{$\downarrow$32.05\%})    
        & 9.36 (\textcolor{black}{$\downarrow$43.78\%})         & 12.07 (\textcolor{black}{$\downarrow$33.04\%})        & 9.86 (\textcolor{black}{$\downarrow$44.67\%}) 
        \\ \midrule
\Ptrnet &\textbf{13.30} (\textcolor{black}{$\downarrow$13.21\%})   &\textbf{9.38} (\textcolor{black}{$\downarrow$52.45\%})    &\textbf{10.94} (\textcolor{black}{$\downarrow$16.29\%})
        &\textbf{13.21} (\textcolor{black}{$\downarrow$17.38\%})   &\textbf{18.07} (\textcolor{black}{$\downarrow$7.73\%})    &\textbf{12.98} (\textcolor{black}{$\downarrow$22.45\%})
        \\

\bottomrule
\end{tabular}
\vspace{\figmargin}
\label{tab:baselines_time}

\end{table*}

\subsubsection{Split by Project}

\Tab~\ref{tab:baselines-split-repo} shows experimental results on datasets split by project. Compared to \Tab~\ref{tab:baselines-multilan}, the performance of all models on all datasets drop consistently, by 26.93\% to 73.41\%. This indicates that the split-by-project scenario is much more difficult than split-by-commit, and models need to have better generalization ability when applied to new projects. 
We also find that on datasets split by project, \Ptrnet shows the best performance.

\begin{table*}[t]
\centering \footnotesize \setlength{\tabcolsep}{5pt}
\caption{The performance on \mcmd split by project under \bleumosesnorm. The score drop rates compared to \Tab~\ref{tab:baselines-multilan} are shown in parentheses.}
\vspace{\figmargin}
\begin{tabular}{>{\bfseries}l *{5}{l} | l} 
\toprule
Model   & \mcmdjava                         & \mcmdcsharp                       & \mcmdcpp                           
        & \mcmdpython                       & \mcmdjs                           & Avg. \\ 
        \midrule
\Commitgen & 5.20 (\textcolor{black}{$\downarrow$58.03\%})     & 4.82 (\textcolor{black}{$\downarrow$73.41\%})       & 4.47 (\textcolor{black}{$\downarrow$61.37\%})      
        & 7.61 (\textcolor{black}{$\downarrow$31.46\%})       & 7.05 (\textcolor{black}{$\downarrow$59.50\%})        & 5.83 (\textcolor{black}{$\downarrow$58.71\%}) 
        \\
\Codisum & 10.23 (\textcolor{black}{$\downarrow$26.93\%}) & 8.43 (\textcolor{black}{$\downarrow$33.78\%}) & 2.87(\textcolor{black}{$\downarrow$76.97\%}) & 9.23(\textcolor{black}{$\downarrow$36.91\%}) & 8.02(\textcolor{black}{$\downarrow$28.65\%})
 & 7.76 (\textcolor{black}{$\downarrow$40.39\%})
        \\ 
\Nmt    & 7.94 (\textcolor{black}{$\downarrow$40.70\%})         & 5.95  (\textcolor{black}{$\downarrow$65.65\%})        & 5.73  (\textcolor{black}{$\downarrow$50.43\%})
        & 5.29 (\textcolor{black}{$\downarrow$54.12\%})         & 7.39  (\textcolor{black}{$\downarrow$56.73\%})        & 6.46  (\textcolor{black}{$\downarrow$54.43\%}) 
        \\
\Nngen  & 5.67 (\textcolor{black}{$\downarrow$68.16\%})         & \textbf{9.89} (\textcolor{black}{$\downarrow$56.85\%})         & 3.90 (\textcolor{black}{$\downarrow$71.51\%})
        & 4.66 (\textcolor{black}{$\downarrow$72.00\%})         & 5.72 (\textcolor{black}{$\downarrow$68.28\%})         & 5.97 (\textcolor{black}{$\downarrow$66.50\%}) \\ 
        \midrule
\Ptrnet &\textbf{7.92} (\textcolor{black}{$\downarrow$48.34\%})    &8.08 (\textcolor{black}{$\downarrow$59.03\%})             &\textbf{6.28} (\textcolor{black}{$\downarrow$51.95\%})
        &\textbf{8.79} (\textcolor{black}{$\downarrow$45.03\%})    &\textbf{11.98}(\textcolor{black}{$\downarrow$38.82\%})    &\textbf{8.61} (\textcolor{black}{$\downarrow$48.57\%}) 
        \\
\bottomrule
\end{tabular}
\vspace{\figmargin}
\label{tab:baselines-split-repo}
\vspace{-10pt}
\end{table*}

Furthermore, to emulate the scenario in which we need to generate commit messages for a new project, we conduct a series of \emph{single project experiments} as described in~\Sec~\ref{sec:rq4} with the NMT model. 
From the results shown in \Tab~\ref{tab:single-project}, we can find that the performance of NMT with Cross-Project training is poor, which is consistent with the previous split-by-project conclusion. Within-Project training is much better than Cross-Project, and 
the performance of NMT model can be further improved through Full-Project training.

\begin{table}[t]
\centering
\footnotesize \setlength{\tabcolsep}{1.5pt}
\caption{\Nmt performance on our dataset \mcmd.} 
\vspace{\figmargin}
\begin{tabular}{>{\bfseries}l *{5}l | l} \toprule
                & \mcmdjava &\mcmdcsharp& \mcmdcpp  &\mcmdpython& \mcmdjs    & Avg.  \\ \midrule
                
Cross-Proj.     & 6.27      & 5.03      & 2.52      & 1.61      & 2.40          & 3.57  \\
Within-Proj.    & 9.35      & 28.82     & \textbf{8.47}      & 7.28      & 18.33         & 14.45 \\ \midrule
Full-Proj.      & \textbf{10.28}     & \textbf{40.04}     & 8.45      & \textbf{7.41}      & \textbf{18.92}         & \textbf{17.02} \\ \bottomrule
\end{tabular}

\label{tab:single-project}
\vspace{-10pt}
\end{table}

\begin{tcolorbox}
\textbf{Summary:}  
The dataset splitting strategies have significant impact on the evaluation of commit message generation models. 
Under the split-by-timestamp or split-by-project strategies, the evaluation scores of models are significantly lower than that of split-by-commit, and the \Ptrnet model is overall the best. 
Moreover, to achieve the best performance, it is recommended to train models with data from both the target project and other projects.

\vspace{-3pt}
\end{tcolorbox}

\section{Discussions}

\subsection{Future Research Directions}

As described in \Sec~\ref{sec:metrics_result}, evaluation metrics are important for the evaluation of commit message generation models, and using different metrics may lead to different conclusions. We have only studied the metrics that have been used in this task. A possible future direction is to study whether other metrics used in some NLP tasks are better than \bleumosesnorm or even design a new metric for commit message generation task.

Another future direction is to leverage more  context information of repositories in the design of commit message generation models. 
In addition to the AST information that \Atom has studied, there are other information that could be explored, such as programming languages' features, contributors, history changes on the target code, associated bug reports, etc. In this regard, our dataset \mcmd, which is information-rich and can be traced back to original repositories to extract all information mentioned above, can be very helpful for future research. We have made our dataset public.

Finally, as discussed in \Sec~\ref{sec:dataset} and \Sec~\ref{sec:splitting}, there are multiple aspects that should attract more attention when evaluating commit message generation models in the future, including multiple PLs, dataset splitting strategies, etc. More specifically, in the settings of split-by-timestamp and split-by-project, the performance of existing models needs to be further improved.

\subsection{Possible Ways to Improve?}

Our findings suggest that commit message generation still has a long way to go. We now show that the performance of commit message generation models could be improved through small changes. Note that we do not aim to design a full-scale model here. Our purpose is to show that there is still ampler room for further improving existing models.

As the experimental results in \Tab~\ref{tab:baselines-multilan} suggest, a retrieval-based model is a simple yet effective approach to generating commit messages. However, \Nngen only uses the ``bag of words'' (which is 1-gram) as the retrieval index, which is basic and can be optimized. 
We can use a simple and different representation. Inspired by the design of BLEU concerning the precision of n-gram matches, we change the representation of diff tokens from 1-gram to n-gram. 
NNGen-Gram4 in \Tab~\ref{tab:retrieval} means that we represent all of the tokens including 1,2,3,4-gram rather than 1-gram only. 
Besides the representation, the retrieval method is also changed in our attempt. As described in~\Sec\ref{sec:metrics_result}, adding a smoothing function can affect the BLEU score. The original BLEU metric used by \Nngen may not be the most suitable option for retrieval. Therefore, we try to add the smoothing function (shown as NNGen-Smooth in~\Tab~\ref{tab:retrieval}) to the retrieval method.
We compare the results of our two attempts NNGen-Gram4 and NNGen-Smooth with \Nngen on \NngenData, as shown in~\Tab\ref{tab:retrieval}. The two variants achieve consistently higher score than the original \Nngen and NNGen-Smooth-Gram4, the combination of them, further improves the performance.

\begin{table}[t]
\centering \footnotesize 
\caption{Retrieval index experiments on \NngenData.}
\vspace{\figmargin}
\begin{tabular}{lllllll}
\toprule
Model         & \bleumoses  & \bleumosesnorm  & \bleunltksmfive \\ 
\midrule
\Nngen         & 16.41           & 23.07       & 16.77 \\ 
\midrule
NNGen-Smooth & 16.47           & 23.12       & 16.80 \\ 
NNGen-Gram4  &   17.26          & 23.92     & 17.60 \\ 
\textbf{NNGen-Smooth-Gram4}  & \textbf{17.48}          &  \textbf{24.03}     & \textbf{17.63} \\ 
\bottomrule
\end{tabular}
\vspace{\figmargin}
\vspace{-2pt}
\label{tab:retrieval}

\end{table}

\subsection{Threats to Validity}

We have identified the main threats to validity as follows:

\begin{itemize}[topsep=0pt,itemsep=0pt,partopsep=0pt,parsep=0pt,leftmargin=8pt]

\item \emph{Data Quality.} The quality of the data could be a threat to validity. 
To mitigate this threat, we have used multiple filtering rules following previous work~\cite{Liu0T0L19} to obtain a set of relatively good-quality commit messages. But there might still exist low-quality data. It is possible to further improve our dataset \mcmd to obtain higher quality.

\item \emph{Human Labeling Bias.} Our manual annotation of the quality of commit message may be biased, and interrater reliability could be a threat to validity: bias may exist in the scores assigned to the same sentence by different raters. We attempted to mitigate this threat by: (1) making clear scoring rules as shown in \Tab~\ref{tab:human_score_meaning} before actual scoring, and (2) having discussions on disagreement cases so that the standard deviations among all 
raters are small. 

\item \emph{Data Sampling.} 100 sampled commit messages are used in the human evaluation.  These samples are randomly selected from the messages with the variance among the three BLEU variants larger than 30, because the aim is to find which BLEU variant correlates with human scores the most, 
especially when there is large variance between the BLEU scores.
This sampling could be a threat to validity. Larger scale human evaluation can be conducted to alleviate this threat.

\item \emph{Programming Languages.} There are plenty of programming languages (PLs) with various characteristics. Although we have expanded the language diversity in \mcmd to 5 popular PLs, it is still not exhaustive. Caution is needed when applying our findings to other PLs.

\item \emph{Replication.} There could exist potential errors in our implementations and experiments. To mitigate this threat, we reuse the existing implementations of the models from the original authors when possible. The new implementation in our experiments (such as data collection, scripts for comparison experiments, and implementation of the improved \Nngen) are double checked by multiple 
experts to ensure correctness. 
\end{itemize}

\section{Conclusion}\label{sec:con}

In this paper, we conduct an in-depth analysis of the datasets and models in the commit message generation task. We have investigated several aspects, including: evaluation metrics, datasets in multiple programming language, and dataset splitting strategies, etc. Our study points out that all these aspects have large impacts on the evaluation. We believe that the results and findings in our study can be of great help for practitioners and researchers working on this interesting area.

Our source code and data are available at \url{https://github.com/DeepSoftwareAnalytics/CommitMsgEmpirical}.

\bibliographystyle{IEEEtran}
\bibliography{ref}

\end{document}